# Superconductivity in Li$_8$Au electride

Xiaohua Zhang,[1] Yansun Yao,[2,*] Shicong Ding,[1] Aitor Bergara[3,4,5], Fei Li,[1] Yong Liu,[1] Xiang-Feng Zhou,[1,6] and Guochun Yang[1,*]

[1]State Key Laboratory of Metastable Materials Science & Technology and Key Laboratory for Microstructural Material Physics of Hebei Province, School of Science, Yanshan University, Qinhuangdao 066004, China;

[2]Department of Physics and Engineering Physics, University of Saskatchewan, Saskatoon, Saskatchewan S7N 5E2, Canada

[3]Departamento de Física, Universidad del País Vasco-Euskal Herriko Unibertsitatea, UPV/EHU, 48080 Bilbao, Spain

[4]Donostia International Physics Center (DIPC), 20018 Donostia, Spain

[5]Centro de Física de Materiales CFM, Centro Mixto CSIC-UPV/EHU, 20018 Donostia, Spain

[6]Center for High Pressure Science (CHiPS), State Key Laboratory of Metastable Materials Science and Technology, Yanshan University, Qinhuangdao 066004, China

E-mail: yansun.yao@usask.ca; yanggc468@nenu.edu.cn

Located at crystal voids, interstitial anion electrons (IAEs) have diverse topologies, which may be tuned to achieve new properties. Elucidating the role of IAEs in electron-phonon coupling (EPC), and using it to design new electride superconductors, leads to the current prediction of superconducting Li$_8$Au at high pressure. We suggest that the occurence of high-temperature superconductivity in electrides requires high-symmetry structures with hydrogen-like cages, an electron acceptor element to balance charges, and isolated IAEs coupled with medium-frequency vibrations. The uniquely designed Li$_8$Au electride has a NaCl-type (B1) lattice, with atomic Au and cubic Li$_8$ cages as bases. Isolated IAEs are formed at the cage centers, with extra charges taken up by Au. These octahedrally coordinated IAEs have a $p$-orbital-like attribute and are strongly coupled with atomic vibrations in the Li$_8$ cages. The strong EPC in Li$_8$Au results in a calculated $T_c$ of 73.1 K at 250 GPa, which is the highest $T_c$ reported to date for all the electrides. A slight substitutional Pt doping can enhance the $T_c$ of Li$_8$Au to exceed liquid nitrogen temperature.

The search for new superconductors is of great interest in condensed matter physics. Over the years, several types of BCS superconductors have been discovered, from earlier A15-type Nb$_3$Ge [1], layer-type MgB$_2$ [2], to recent perovskite-type H$_3$S [3], and sodalite-type LaH$_{10}$ [4,5]. The discovered $T_c$ in BCS superconductors grows continuously, from the early thought upper limit around 40 K ('McMillian limit') to above 260 K in recently discovered hydrides [5,6]. Interestingly, high-$T_c$ hydrides tend to have high symmetry structures, which has played a guiding role in the search for new superconductors [7,8].

The co-occurrence of superconducting and electride states are less explored. In electrides, a fraction of electrons breaks away from atoms to reside in the interstitial voids, behaving like nucleus-free anions (IAEs) [9], which can be tuned in magnitude and topology [10-12] to achieve new properties. Alkali metals become electrides under high pressure [13,14] and exhibit moderate superconductivity ($T_c$ < 20 K)

[15,16]. Incorporating non-mental elements into the electrides can modulate the IAE and enhance superconductivity. For example, Li$_5$C [17], Li$_5$N [18], and Li$_6$P [19] present interconnected IAEs which enhance the $T_c$ to 48.3 K, 48.97 K, and 39.3 K, respectively.

High-$T_c$ hydrides tend to favor high-symmetry configurations of hydrogens (H), e.g., H$_8$ [20], H$_{24}$ [21], and H$_{32}$ [22] cages. One approach to turn these cages into high-symmetry electrides is replacing H's by elements with same valence configuration but more electropositive [23,24]. Lithium (Li) is an obvious option, since it behaves like H at high pressure and maintains electride states that favors superconductivity (Fig. 1) [15,25]. In addition, Li has the ability to form clusters of various stoichiometries ('superatoms') when alloyed with some transition metals [26]. The formation of electrides generally requires that Li has a higher stoichiometric ratio than the other metal [27]. Therefore, the stabilization of high symmetry electrides of Li requires selecting an electron acceptor to satisfy geometry and charge balance. Among the elements, gold (Au) stands out due to its unique chemical attributes and the ability to present high oxidation states at high pressure (Fig. 1) [28-30]. As a host, Au also presents a possibility of stabilizing Li-Au electrides [31].

In this Letter, we report a novel $Li_8Au$ electride with good superconducvitiy. The structure was inspired by previously reported $Li_3Au$, in which Au and edge shared *bcc* $Li_3$ occupy two *fcc* sites in a NaCl lattice (Fig. 2b). In $Li_8Au$, we kept the NaCl motif, but replaced the $Li_3$ with cubic $Li_8$ cages, isostructural to the $H_8$ cages [20, 32-34]. IAEs are formed at the centers of the $Li_8$ cages, and arranged in another *fcc* sublattice. The strong electron-phonon coupling (EPC) in $Li_8Au$ induces a high $T_c$ of ~73.1 K at 250 GPa, much higher than in previously reported electrides.

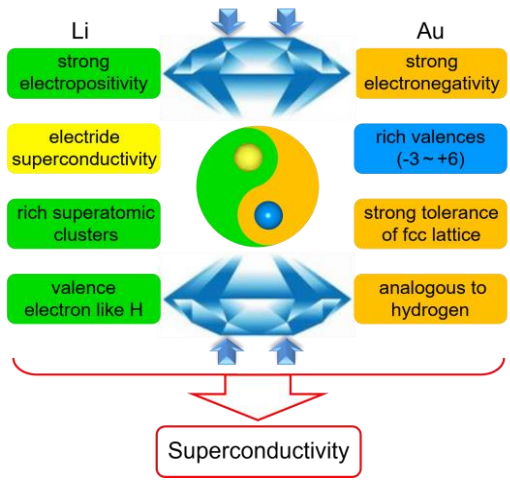

**FIG. 1.** Schematic illustration of the design strategy for superconducting electrides consisting of Li and Au. The special features of Li and Au that occur under high pressure are highlighted in yellow and blue, respectively.

Hydrogen cages are important intermediate structures between molecular $H_2$ and atomic H, which, after being precompressed by metal atoms, contributes enormously to the EPC. Simple-cubic (*sc*) $H_8$ is a common H cage [20, 22] observed in hydrides (Fig. 2a). Considering the similarity between Li and H, the formation of cubic $Li_8$ cage should be possible, but it would require weaker bonds, which means that some valence electrons need to be removed from the cage. The cubic voids naturally accommodate the extra electrons, forming IAEs (Fig. 2a). It should be noted that crystalline counterpart of simple cubic is dynamically unstable, and even less so at high pressures when Li departs from nearly free electron behavior (Fig. S1). At high pressure, the reduced interstitial sites in crystalline simple cubic could not store sufficient IAEs required to stabilize the structure, and therefore elemental Li must adopt other structures [15]. Au is a suitable candidate for additional 'electron reservoir' in the crystal due to its exceptional charge adjustability, which can act as both an electron donor and acceptor [28,31]. Furthermore, the *fcc* sublattice of Au is robust to neighboring superatoms (Fig. S2), ideal for hosting *sc* $Li_8$ in the crystal.

We replace the edge-sharing *bcc* units in $Li_3Au$ with IAEs-centered $Li_8$ units. This results in a stoichiometric $Li_8Au$ within the same *Fm*-3*m* space group (Fig. 2c). The IAE has a smaller volume than Li atom (Table. S1), which means that at high pressure the IAE-centered cubic unit is energetically more favorable than its atom-centered counterpart. Indeed, the convex hull of the Li-Au phase diagram indicates that $Li_8Au$ will replace $Li_3Au$ to become thermodynamically stable at high pressure (above 177.3 GPa) (Fig. S3). Furthermore, *Fm*-3*m* $Li_8Au$ becomes dynamically stable in the pressure range of 120-300 GPa (Fig. S4) as confirmed by phonon calculations. All computational details can be found in the Supplemental Material [35].

The stability of $Li_8Au$ is enhanced by charge-transfer-induced ionic interaction. The $Li_8$ cage loses a total charge of 5.24 e$^-$ to the surroundings. The two electron acceptors, the IAE and Au atom, gain 0.74 e$^-$ and 4.50 e$^-$ (Fig. 2c), respectively. Thus, the IAE-centered $Li_8$ is positively charged, behaving as a superatomic cation. The electrons transferred to Au atom must populate its 6*p* orbitals (details below), which causes the expansion of its atomic radius, allowing a larger space to accommodate IAE-centered $Li_8$. In addition, the Li-Li bond interaction in IAE-centered $Li_8$ (Fig. S5), induced by electron transfer, also contribute to the stability of $Li_8Au$.

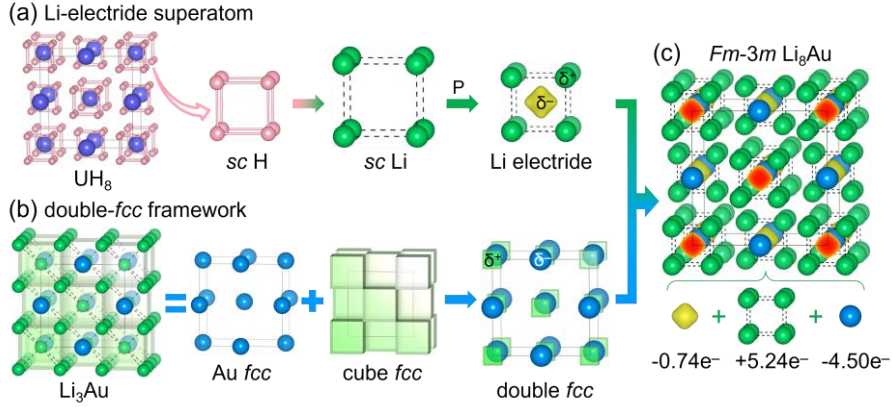

**FIG. 2.** (a) Cubic Li electride superatom analogous to the H$_8$ cage in UH$_8$. (b) The NaCl lattice in Li$_3$Au consisting of Li$_3$ and Au bases. (c) Crystal structure of *Fm*-3*m* Li$_8$Au. The IAEs are shown by yellow spots. Numbers below identities are the calculated Bader charges of their constitutional units at 200 GPa (+/– indicates electronic lose/gain).

Li$_8$Au presents regular IAEs in a high-symmetry metallic structure. The electronic band structure shows three bands crossing the Fermi level ($E_F$), mainly contributed by Li 2$p$, Au 6$p$, and IAEs (Fig. 3a). These bands feature a simultaneous occurrence of flat bands (W-L, X-W-K paths) and steep bands (W-L, X-W-K paths), which sets up a favorable condition for electron pairing. The three corresponding Fermi surfaces (FSs) are shown in Fig. 3b. Both FS 1 and 2 are derived mainly from a mixture of Li 2$p$ and Au 6$p$ states, while FS 3 primarily from the hybridized states of Li 2$p$ and IAEs, and hybridized states of Li 2$p$ and Au 6$p$. This indicates that Li 2$p$ acts as a common channel interacting with all other states. Based on the orbital hybridization, IAE's octahedral topology, and the orbital symmetry matching rule, we infer that IAEs in Li$_8$Au show a $p$-orbital-like attribute (Fig. S6). The maximum Fermi velocity is concentrated at two centrosymmetric points on the FS 2, whereas the medium and low Fermi velocities have a symmetrical distribution on all three FSs (Fig. 3b), indicating strong FS nesting [54]. The nesting function $\xi(Q)$ shows that a considerable region of the FS is nested by Γ-X-W-K vectors (Fig. 3c) [55]. Moreover, a sharp van Hove singularity (vHs) occurs at -1.09 eV below the $E_F$ (Fig. S7), similar to H$_3$S [56,57] and LaH$_{10}$ [58]. These electronic structures have been shown to be favorable for the formation of stable Cooper pairs [17,59].

The superconductivity of Li$_8$Au is evaluated via EPC calculations and the Allen-Dynes modified McMillian formula [60,61]. As shown in Fig. 3d, the EPC constant $\lambda$ increases notably between 150 and 270 GPa, mainly due to the softening of acoustic branches around the L point and the low-energy optical branches around L and the Γ-X path (Figs. 3e and S8). However, the phonon frequency logarithmic average $\omega_{\log}$ decreases with pressure. The two competing mechanisms result in a $T_c$ peaking at 250 GPa. Using a Coulomb pseudopotential of $\mu^* = 0.1$, the calculated $T_c$ at 250 GPa is 68.5 K (Fig. 3d and Table S2), and considering its semiempirical character [62], the estimate of $T_c$ when $\mu^*$ is in a 0.08-0.13 range goes between 73.1 and 62.2 K (Fig. S9). With inclusion of spin-orbital coupling, the $T_c$ of Li$_8$Au is calculated to be 66.3 K with $\mu^* = 0.1$ (Fig. S10). All these $T_c$ values exceed the highest $T_c$ reported to date for electrides (48.3 K for Li$_5$C [17], 48.97 K for Li$_5$N [18] and others (Fig. S11)), as well as higher than the $T_c$'s of Au compounds (~30 K for Ba(AuH$_2$)$_2$ [63]). Moreover, Li$_8$Au is predicted to be a single-gap superconductor, corresponding to a $T_c$ of 81.8 K (Fig. S12) based on electron-phonon Wannier calculations [64]. Given a sharp DOS peak (vHs) below the $E_F$ in Li$_8$Au, replacing a small amount of Au with Pt, acting as a hole donor, might boost $N_{E_F}$ to yield higher $T_c$. Using virtual crystal approximation [65], an optimal $T_c$ value of 78.3 K (with $\mu^* = 0.1$) is achieved at Pt doping close to 0.5% (Figs. 3f and S13-S15). For superconductors with light mass element such as hydrides, nuclear quantum effect (NQE) may have effects on the $T_c$ [66-68]. However, NQE has been shown to have neglectable effects on superconducting Li [69] and therefore not included in the superconductivity study of Li$_8$Au.

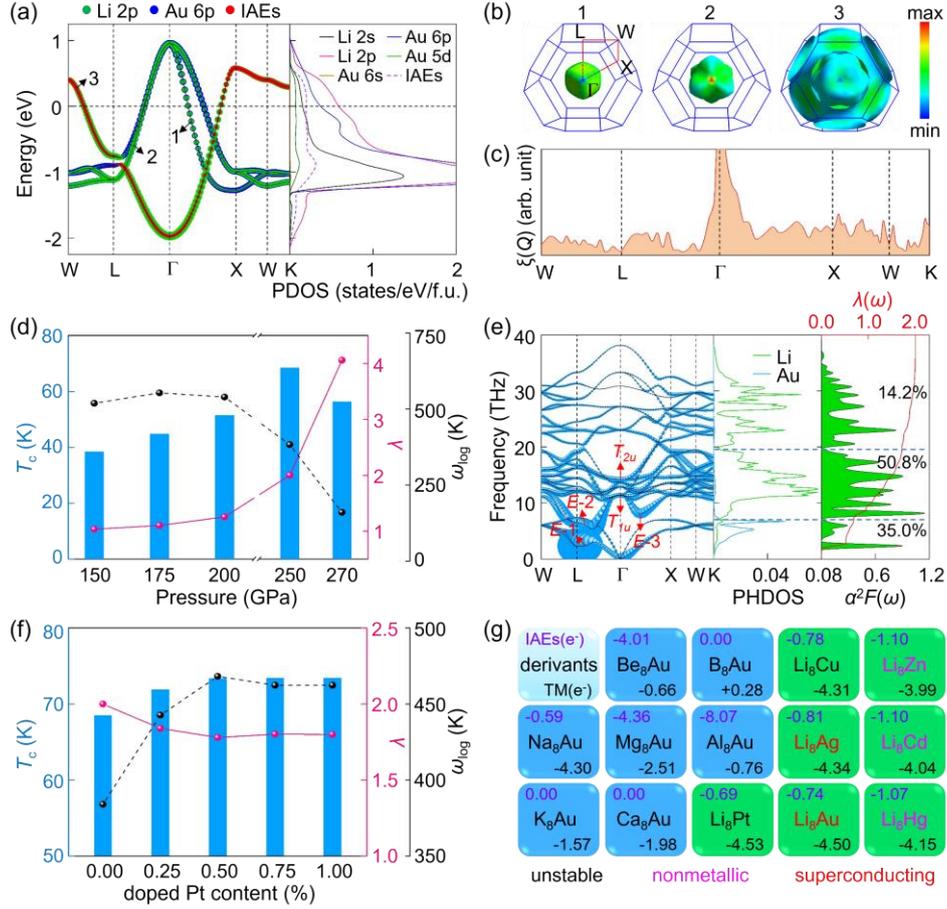

**FIG. 3.** (a) Orbital-projected electronic bands and density of states (DOS). (b) Fermi surfaces corresponding to the three bands crossing the $E_F$, color-coded by the Fermi velocity. (c) The nesting function $\xi(Q)$ along high-symmetry paths. (d) Calculated $\lambda$, $\omega_{\log}$, and $T_c$ of Li$_8$Au at different pressures. (e) Phonon dispersion relations, projected phonon density of states (PHDOS), Eliashberg spectral function $\alpha^2F(\omega)$ and frequency-dependent EPC parameter $\lambda(\omega)$. The size of solid dots on the phonon spectra signifies the contribution to the EPC ($\lambda_{q,v}$). (f) The evolution of $\lambda$, $\omega_{\log}$, and $T_c$ of Pt-doped Li$_8$Au with increasing Pt content at 250 GPa. (g) Derivatives of Li$_8$Au substituting Li or Au with their neighboring elements. Dynamically unstable structures are marked in black and stable ones in color. The values represent the calculated Bader charges of IAEs (purple) and En atoms (black) at 200 GPa.

The PHDOS and Eliashberg spectral function $\alpha^2F(\omega)$ can be divided into three regions, low-frequency region (below 7 THz) dominated by Au atom, Li$_8$ cage-derived intermediate-frequency region (7-20 THz) and high-frequency (above 20 THz) regions (Fig. 3e), which make the contribution of 35.0%, 50.8%, and 14.2% to total $\lambda$, respectively. The strongest local EPC are from two softened acoustic modes around the L point (Fig. 3e), *i.e.*, the twofold degenerate $E$-1 and $E$-2 modes. These modes represent stretching vibrations of pairs of Li atoms across the body diagonal in Li$_8$ cage, which favors a strong extrusion interaction with IAEs (Fig. S16). Meanwhile, Au atom also participates in the vibrations, exhibiting a low-frequency feature. These two modes induce a shift in energy around the L point, and a removal of degeneracy around the Γ point (Fig. S17). The mechanism by which the $E$-1 and $E$-2 modes promote the EPC is similar to that of the $E_{2g}$ mode in MgB$_2$ [70]. In the intermediate-frequency region, the three modes ($T_{1u}$, $T_{2u}$, and soft $E$-3) associated with twisting vibrations of Li$_8$ cage (Fig. S18) induce a minor shift of the flat bands along W-L and X-W-K paths (Fig.

S19). Therefore, the Li$_8$ cage and IAEs dominate superconductivity in Li$_8$Au.

Finally, we explore the key factors for stabilizing this electride using elemental substitutions for Li and Au (Fig. 3g). For clarity, the electropositive atoms (replacing Li) are denoted by Ep, and the electronegative ones (replacing Au) are En. It turns out that no Ep$_8$Au, apart from Li$_8$Au, are dynamically stable (Figs. S20-21, Fig. 3g). Li$_8$En, on the other hand, has several dynamical stable structures, e.g., Li$_8$Ag, Li$_8$Zn, Li$_8$Cd, and Li$_8$Hg, which shows robustness to electronegative replacement (Figs. 3g and S22-23). Interestingly, Li$_8$Ag is metallic (Figs. S24-25), while Li$_8$Zn, Li$_8$Cd, and Li$_8$Hg are all semiconducting (Fig. S26). This suggests that electronic properties in double-*fcc* electride can be further tuned through isotypic replacement or doping.

In summary, we have designed a new electride material, Li$_8$Au, consisting of a double *fcc* lattice with atomic Au and cubic Li$_8$ cages as bases. The IAEs are located at the cage centers, forming octahedron-like IAEs with a *p*-orbital-type attribute. This topology of the IAEs induces strong coupling to the vibration of Li$_8$ cages, and enhances the phonon mediated superconductivity. Li$_8$Au is calculated to have the highest superconducting $T_c$ among all reported electrides to date. Moreover, the $T_c$ of Li$_8$Au can be further increased by substitutional Pt doping. Our work serves as a guidance to design superconducting electrides with high-symmetry building blocks.

## Acknowledgments

This work was supported by the National Key Research and Development Program of China (Grant No. 2022YFA1402300), the Natural Science Foundation of China under Grants No. 52025026, No. 21873017, and No. 21573037, the Postdoctoral Science Foundation of China under Grant No. 2013M541283, the Innovation Capability Improvement Project of Hebei province (Grant No. 22567605H), the Natural Science Foundation of Hebei Province (Grant No. B2021203030), and Natural Sciences and Engineering Research Council of Canada (NSERC). A.B. acknowledges financial support from the Spanish Ministry of Science and Innovation (Grant No. PID2019-105488GB-I00) and the Department of Education, Universities and Research of the Basque Government and the University of the Basque Country (IT1707-22).